\documentclass[twocolumn, prl]{revtex4}
\usepackage{booktabs}
\usepackage{graphicx}
\usepackage{dcolumn}
\usepackage{bm}
\usepackage{amsmath}

\usepackage{color}

\def\nar{\ref@jnl{New A Rev.}}          

\def\msun{M_\odot}

\begin{document}

\title[The Loudest Events]{The Loudest Gravitational Wave Events}

\author{Hsin-Yu Chen$^1$ and Daniel E. Holz$^2$}
\affiliation{$^1$Department of Astronomy and Astrophysics, University of
  Chicago, Chicago, IL 60637\\
$^2$Enrico Fermi Institute, Department of Physics, and Kavli Institute
for Cosmological Physics\\University of Chicago, Chicago, IL 60637}


\begin{abstract}
As first emphasized by Bernard Schutz, there exists a universal distribution of signal-to-noise ratios for gravitational
wave detection.
Because gravitational waves (GWs) are almost impossible to obscure via dust absorption or other
astrophysical processes, the strength of the detected signal is dictated solely by the
emission strength and the distance to the source.
Assuming that the space
density of an arbitrary population of GW sources does not evolve, we show
explicitly that the
distribution of detected signal-to-noise (SNR) values depends solely on the
detection threshold; it is independent of the detector network (interferometer
or pulsar timing array), the individual detector noise curves (initial or
Advanced LIGO), the nature of the GW sources (compact binary coalescence,
supernova, or some other discrete source), and the distributions of source
variables such as the binary masses and spins (only non-spinning neutron stars
of mass exactly $1.4\,\msun$ or a complicated distribution of masses and
spins).  We derive the SNR distribution for each individual detector
within a network as a function of the relative detector orientations and
sensitivities.  While most detections will have SNR near the detection
threshold, there will be a tail of events to higher SNR. We derive the SNR
distribution of the loudest (highest SNR) events in any given sample of
detections.
We find
that in 50\% of cases the loudest event out of the first four should have an SNR
louder than 22 (for a threshold of 12, appropriate for the Advanced LIGO/Virgo
network), increasing to a loudest SNR of 47 for 40 detections.
We expect
these loudest events to provide particularly powerful constraints on their source parameters,
and they will play an important role in extracting astrophysics from
gravitational wave sources.  These distributions also offer an important
internal calibration of the response of the GW detector networks.
\end{abstract}

\maketitle

\section{\label{sec:intro}Introduction}

Gravitational waves (GWs) couple very weakly to matter. The downside of this
is that they are difficult to detect, and almost a century after
they were first predicted by Einstein~\citep{1916SPAW.......688E} they remain to be directly detected on Earth. The
upside is that GWs propagate with little interference, being almost impossible
to absorb or scatter, and thus cleanly carry information from the source to the
observer. As a result both the amplitude and the measured signal-to-noise (SNR) ratio
of GWs scale inversely with luminosity distance, 
leading to a universal SNR distribution of GW events as a
function of the SNR detection threshold~\cite{2011CQGra..28l5023S}.
This follows directly from the simple relationship betwen distance
and volume, and applies so long as the source population does not evolve with
distance.


In the first several years of Advanced LIGO/Virgo operations we expect to have
tens of detections per year of gravitational waves from compact binary
coalescence~\cite{2010CQGra..27q3001A,2013arXiv1304.0670L,2013PhRvL.111r1101C}. 
These detections must follow the universal SNR distribution, and
  this offers an important internal self-calibration of the GW detector
  network. Additionally, the distribution offers a simple internal test of whether the
  first events are statistically consistent with expectations.
Although most of these detections will be found with SNRs close
to the detection threshold, there will exist a tail to higher SNR.
Fisher matrix calculations show
that the timing, chirp mass, and amplitude measurement all improve as
$\sim1/\mbox{SNR}$~\cite{1994PhRvD..49.2658C,2013PhRvD..88f2001A}. The
highest SNR events will likely offer the best constraints on
both intrinsic and extrinsic parameters of their sources, and thereby
enable important physics and 
astrophysics~\cite{2013PhRvD..88f2001A}.
For example, accurate determination of binary masses helps distinguish between neutron stars
and stellar mass black holes, and elucidates the ``mass
gap'' problem~\cite{1998ApJ...499..367B,2012ApJ...757...91B}. Higher SNR measurement of waveforms
may help probe the neutron star
equation-of-state~\cite{2012PhRvL.108a1101B,2013ApJ...773...11H,PhysRevLett.113.091104}.
These loudest events are likely to have improved sky localization, increasing the
probability of observing electromagnetic (EM) counterparts to the GW events and
leading to the birth of GW/EM multi-messenger
astronomy~\cite{2012ApJ...746...48M}. In particular, joint detections would
confirm binary systems as the progenitors of short-hard gamma-ray
bursts~\cite{2012ApJ...760...12A,2011NewAR..55....1B}, probe the Hubble
constant, and potentially measure the dark energy equation of
state~\cite{1986Natur.323..310S,2005ApJ...629...15H,2006PhRvD..74f3006D,2013arXiv1307.2638N}.
We argue that these loudest
events {\em must}\/ exist, and will play an important role in the coming age of
gravitational-wave astrophysics.




\section{\label{sec:SNR}The Universal SNR Distribution}

A given GW network will detect some number of GW events, with each event
characterized by a measured signal-to-noise ratio (SNR), $\rho$. We are interested in the
distribution of $\rho$. We assume that the space density and intrinsic
properties of the source population do not evolve. This is justified given that
the Advanced LIGO/Virgo network is
only able to probe the nearby universe,
$z\lesssim0.2$~\cite{2013arXiv1304.0670L} (although see~\cite{2014ApJ...789..120B} for an
  example where this is not the case).
For the sake of definiteness and to enable Monte
Carlo comparisons, in what follows we will
assume that the GW sources are merging compact binaries, although our results
are independent of this assumption and are valid for any discrete distribution
of sources.
Following ~\cite{1987thyg.book..330T,1994PhRvD..49.2658C,PhysRevD.53.2878}, 
we compute the SNR of a binary inspiral and merger assuming a restricted post-Newtonian waveform observed by 
a network of ground-based GW detectors:
\begin{equation} \label{eq:snr}
\rho^2=4\frac{{\cal A}^2}{D_L^2}[F_{+}^2(\theta,\phi,\psi)(1+{\rm cos}^2\,\iota)^2+4F_{\times}^2(\theta,\phi,\psi) {\rm cos}^2\,\iota]I_7,
\end{equation}
where ${\cal A}=\sqrt{5/96}\pi^{-2/3}(G{\cal M}_z/c^3)^{5/6}c$, 
${\cal
  M}_z=(1+z)(m_1m_2)^{3/5}/(m_1+m_2)^{1/5}$ is the redshifted chirp mass, $D_L$
is the luminosity distance, 
$\psi$ is the orientation of the
binary within the plane of sky,
and $F_+$ and $F_{\times}$ are the detector antenna power patterns,  which are themselves
functions of the source sky location, $(\theta,\phi)$.
The inclination angle between the binary's rotation axis and the line of
sight is given by $\iota$. The noise curve of the detector is encapsulated in
$I_7$, which is an integral
over the detector's power spectral density $S_h(f)$: 
$$I_7=\displaystyle\int_{f_{\rm low}}^{f_{\rm high}} \frac{f^{-7/3}}{S_h(f)}\,df.$$

\begin{figure}
\centering 
\includegraphics[width=0.99\columnwidth]{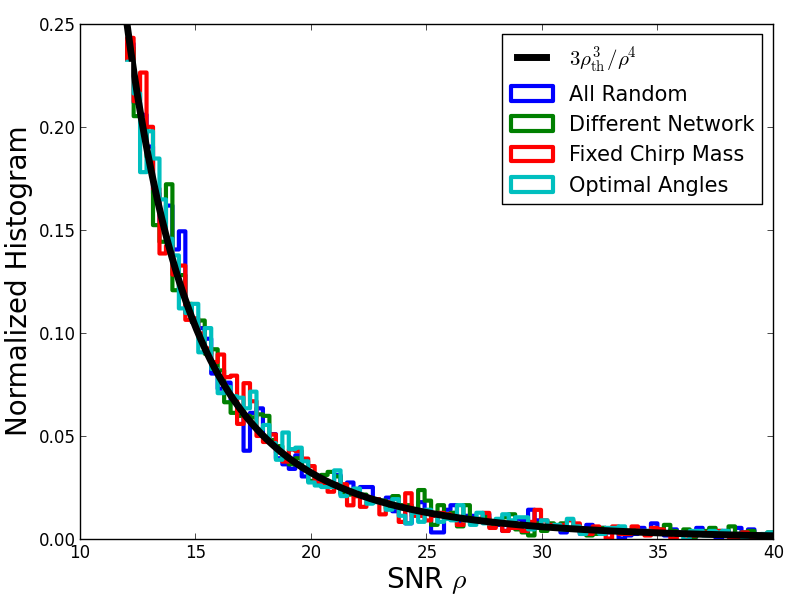}
\caption{\label{fig:distribution}
Universal distribution of SNR (Eq.~\ref{eq:distribution}), for $\rho_{\rm th}=12$, plotted
as the solid black curve. 
The colored histograms show the results from our Monte Carlo simulations of 5,000 detections. ``All
Random'' is from the basic simulation described in the text, and verifies our
analytic prediction. 
The results are independent of the network properties (``Different
Network'' has a different number of detectors, with different relative orientations
and noise curves), the chirp mass distribution (``Fixed Chirp 
Mass''), and the sky location and inclination distribution (``Optimal Angles''
has all binaries ``overhead'' and face-on). This shows
explicitly that the SNR distribution is universal. 
}
\end{figure}


The probability of a merger lying within an
infinitesimal interval $dD$ near the comoving distance $D$ is given by $f_DdD
\propto D^2 dD$, where $f_D=f(D)$ is the distribution of the source
comoving distances~\footnote{This is only true at low redshifts ($z\lesssim0.5$), appropriate for
  stellar mass compact binary sources detected by upcoming advanced ground-based
  detector networks. At high redshift
  the full comoving volume needs to be incorporated.}.
In the nearby universe we can approximate the luminosity distance by the comoving distance. 
We see from Eq.~\ref{eq:snr} that the SNR scales as 
$\rho \propto 1/{D_{\rm L}}$; we note that this is true for {\em all}\/ GW
sources, not just for binary
coalescences~(e.g.,~\cite{1987thyg.book..330T}). 
The other extrinsic parameters (sky
location, binary orientation, and inclination) are randomly distributed and
do not impact the final distributions, as shown explicitly below. If we assume
that the chirp mass distribution and the space density do not evolve with
distance, then the resulting distribution of SNR only depends upon the distance. We find
$$ 
f_{\rho}=f_{D_L}\left|\frac{d D_L}{d\rho} \right|=f_{D}\left|\frac{d D}{d\rho} \right|\propto \frac{1}{\rho^2} \frac{1}{\rho^2}=\frac{1}{\rho^4}.
$$
where the second equality is only true at low redshift. 
Normalizing this for a given network SNR threshold, $\rho_{\rm th}$, we find
that the distribution of SNRs for sources in the local universe is exactly described by
\begin{equation} \label{eq:distribution}
f_{\rho}=\frac{3\rho_{\rm th}^3}{\rho^4}.
\end{equation}
This is identical to Eq.~24 of~\cite{2011CQGra..28l5023S}.

In Fig.~\ref{fig:distribution} we plot this distribution assuming $\rho_{\rm
  th}=12$. The distribution peaks at the threshold value, and has a tail to
  higher SNR events. For explicit comparison we have also performed Monte Carlo
simulations of the detection of a binary population, sampling over the full parameter space
$(D_L,\theta,\phi,\psi,\iota)$ with random sky locations and binary
orientations, and with the total mass of the binaries, $M_{\rm tot}=m_1+m_2$,
drawn uniformly between $2\msun$ and $20\msun$ and $m_1$ 
drawn uniformly between $1\msun$ and $M_{\rm tot}-1\msun$. For each randomly drawn binary we use Eq.~\ref{eq:snr} to
calculate the SNR of the simulated events for a given GW network. As shown in 
Fig.~\ref{fig:distribution}, the histograms
of SNR for our various simulated populations follow our predictions. 
The distribution of SNR presented in Eq.~\ref{eq:distribution} is
  universal, and is what will be found in all GW detectors for all non-evolving, low-redshift GW source
populations.


\section{\label{sec:loudest}The Universal Loudest Event Distribution}

We now turn our attention to the high-SNR tail of events, and make
predictions for the highest SNR event out of any
detected sample of GW events.
We assume that within any arbitrary GW network we have $N$
compact binary detections
with SNR values given by $\{\rho_1,\rho_2,...,\rho_N\}$. We define the loudest 
event as ${\rho_{\rm 
    max}}={\rm max}\{\rho_1,\rho_2,...,\rho_N \}$.
The probability of
$\rho_{\rm max}$ being less than a given value $\rho$ is
\begin{align*}
{P(\rho_{\rm max} < \rho)}&={P(\rho_1<\rho; \rho_2<\rho;...;\rho_N<\rho)}\\
&=P(\rho_1<\rho)P(\rho_2<\rho)...P(\rho_N<\rho)\\
&=(F_{\rho})^N,
\end{align*}
where the second line follows from the assumption that each event is
independent. $F_{\rho}$ is the cumulative distribution function of
$\rho$ and can be computed by integrating Eq.~\ref{eq:distribution} from
$\rho_{\rm th}$ to any desired value of $\rho$.
$P(\rho_{\rm max} < \rho)$ is equivalent to the cumulative distribution function, $F_{\rho_{\rm max}}$, of the loudest event, 
$\rho_{\rm max}$. The
probability distribution function of the loudest event,
$f_{\rho_{\rm max}}$, is obtained by taking a derivative:
\begin{align} \label{eq:maxsnr}
\nonumber f_{\rho_{\rm max}}&=\frac{dF_{\rho_{\rm max}}}{d\rho_{\rm max}}=\frac{d(F_{\rho})^N}{d\rho}\bigg|_{\rho=\rho_{\rm max}}	\\
&=\frac{3N}{\rho_{\rm max}}\left(\frac{\rho_{\rm th}}{\rho_{\rm max}} \right)^3 \left[1-\left(\frac{\rho_{\rm th}}{\rho_{\rm max}} \right)^3 \right]^{N-1}.
\end{align}
We have verified this distribution explicitly using Monte Carlo techniques. 
See~\cite{2012PhRvD..86h2001V} for an alternative approach to deriving this distribution.

\begin{figure}
\centering 
\includegraphics[width=0.99\columnwidth]{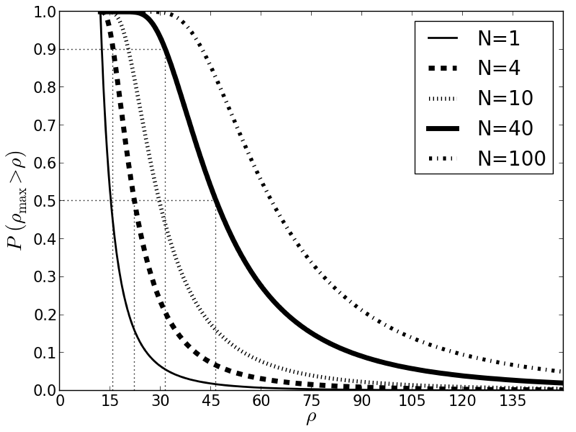}
\caption{\label{fig:plarge}
Probability that the loudest event of a given sample of detections has SNR greater than $\rho$.
We take the network threshold to be $\rho_{\rm th}=12$, and we assume samples
sizes of $N=1$, 4, 10, 40, and 100 detections. We find that in 90$\%$ of
cases the loudest event among the first four detections ($N=4$) will be louder
than $\rho_{\rm max}=15.8$, while 50$\%$ of cases will have a loudest event with
$\rho_{\rm max}>22.2$.
For a sample of 40 detections these rise to $\rho_{\rm max}>31$ (90$\%$) and
$\rho_{\rm max}>47$ (50$\%$). Notice that the $N=4$ curve is visibly shifted
from the $N=1$ case, implying that the loudest event out of a small number of
detections can be significantly louder than a typical ``threshold'' event.
}
\end{figure}

We are now able to forecast the distribution of the loudest events, as shown in Fig.~\ref{fig:plarge}. The
probability that the loudest event is louder than $\rho$, out of $N$ detections above a detection
threshold $\rho_{\rm th}$, is given by $P(\rho_{\rm max}>\rho)$, and
can be calculated by integrating Eq.~\ref{eq:maxsnr} from $\rho$ to infinity.
For example, if we set the network detection
threshold to $\rho_{\rm th}=12$ (appropriate for the case of Advanced LIGO/Virgo) we find that 90$\%$ 
of the time the
loudest event out of the first 4 detections will have $\rho_{\rm
  max}>15.8$. The loudest event out of the first 40 detections, corresponding
roughly to one year of observation with Advanced
LIGO/Virgo~\cite{2010CQGra..27q3001A,2013arXiv1304.0670L,2013PhRvL.111r1101C},
will have $\rho_{\rm max}>31$. Half the time we will find the loudest event to have
$\rho>22$ for 4 events and $\rho>47$ for 40 events. We emphasize that these
statements are independent of the specific noise curves or configurations of the
detector network or even the nature of the source population.

\begin{figure}
\centering 
\includegraphics[width=0.99\columnwidth]{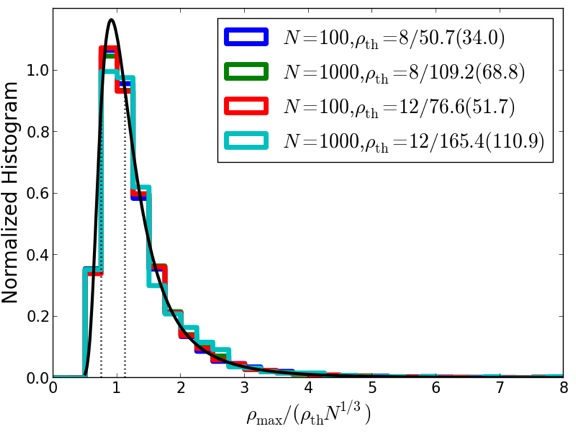}
\caption{\label{fig:normalized} Rescaled histogram of $\rho_{\rm max}/(\rho_{\rm
    th} N^{1/3})$ from $N=100$ and $N=1,000$ detections,
  repeating each random sample ten thousand times, and with $\rho_{\rm th}$ taken to be 8 or
  12. The average and standard deviation of $\rho_{\rm max}$ before rescaling
  are shown in the legend for reference.  In all cases the distributions of
  $\rho_{\rm max}$ follow a similar shape after rescaling. For
  90$\%$ of the cases (left vertical dotted line), $\rho_{\rm max}>0.76\rho_{\rm th} N^{1/3}$, while for 50$\%$ of
  the cases (right dotted line), $\rho_{\rm max}>1.13\rho_{\rm th} N^{1/3}$. }
\end{figure}

The distribution of the loudest events given in Eq.~\ref{eq:maxsnr} depends upon only two
parameters: the number of detections, $N$, and the detection threshold,
$\rho_{\rm th}$. Since the shape of the distribution is similar for all $N$ and
$\rho_{\rm th}$, we are able to find a scaling to produce a universal distribution.
We define a new variable, $y\equiv\rho_{\rm max}/a$, where $a$ is an arbitrary
scaling. The distribution of $y$ values becomes
$$f_{y}=\frac{d \rho_{\rm max}}{dy}f_{\rho_{\rm max}}=a\frac{3N}{ay}\frac{\rho_{\rm th}^3}{a^3y^3}\left[ 1-\frac{\rho_{\rm th}^3}{a^3 y^3} \right]^{N-1}.$$
If we set $a=\rho_{\rm th}N^{1/3}$ this distribution becomes
$$f_{y}=\frac{3}{y^4}\left[ 1-\frac{1}{Ny^3} \right].$$
We note that this distribution is independent of $\rho_{\rm th}$ and $N$ when
$N$ is large. In Fig.~\ref{fig:normalized} we explicitly show that this scaling
produces a universal
form for the distribution of the loudest events.
Furthermore, from this distribution we are able to produce generic, simple, and powerful
statistical predictions. For example, we conclude that in 90$\%$ of cases,
$\rho_{\rm max}>0.76\rho_{\rm  
  th} N^{1/3}$, while in 50$\%$ of cases, $\rho_{\rm max}>1.13\rho_{\rm th}
N^{1/3}$.  Comparing to the exact analytic form, these
expressions are good to 8$\%$ for 4 detections and 0.6$\%$
for 40 detections.

\section{The SNR Distribution in Individual Detectors}
\begin{figure}[t]
\centering 
\includegraphics[width=0.99\columnwidth]{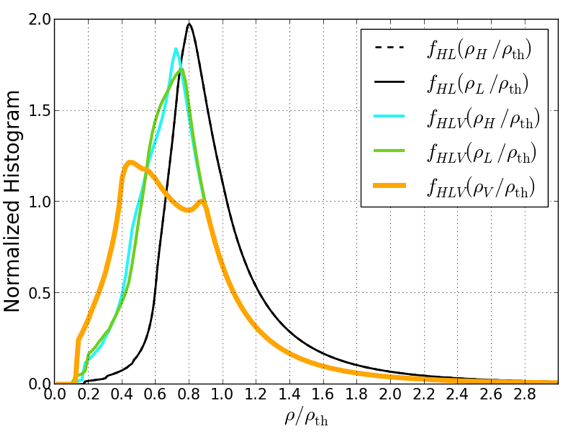}

\caption{\label{fig:scale}
Scaled SNR distributions for individual detectors as part of an HL or HLV
network, for identical detector sensitivities. These distributions are
universal and can be applied for any network threshold. Note in particular that
the Virgo detector finds lower SNR values than the LIGO detectors, even when all are operating at equivalent
sensitivity.
}
\end{figure}

In Eq.~\ref{eq:distribution} we show the distribution of SNR for an arbitrary
detector network, where the network SNR is the root of the sum of the squares of the
SNRs in each individual detector comprising the network. In addition to the overall source amplitude,
the signal strength in each detector depends on the individual detector's
sensitivity and the relative orientation between the source and each
detector. For a detector network with a given network threshold, the distribution
of SNR in each {\em individual}\/ detector can be calculated:
\begin{equation}\label{eq:indsnr}
f_{\rho_i}=\displaystyle\int_{\rho_i \ge \rho_{\rm eff}} f(\rho_i|\theta,\phi,\psi,\iota)\, f_{\rm det}(\theta,\phi,\psi,\iota)\, d\theta d\phi d\psi d\iota
\end{equation}
where $f(\rho_i|\theta,\phi,\psi,\iota) = 3 \rho_{\rm eff}^3 /\rho_i^4 $ and
$$
\rho_{\rm eff}\equiv {\rho_{\rm th}}/{\left ( \displaystyle \sum_{j=1}^N
  R_{ji}\right )^{1/2}}$$
with $R_{ji}\equiv {\rho_j^2}/{\rho_i^2}$.
The prior on sky location and binary orientation depends upon the antenna power
pattern, $P_j$~\cite{2011CQGra..28l5023S}:
$$
f_{\rm det}(\theta,\phi,\psi,\iota)=\frac{1}{n} \left( \displaystyle \sum_{j=1}^{N} P_j \right )^{3/2}\; {\rm sin}\,\theta \; {\rm sin}\,\iota,
$$
where the normalization factor $n$ is integrated over $d\theta\,d\phi\,d\psi\,d\iota$.
We simplify Eq.~\ref{eq:indsnr} by rewriting the individual SNR, $\rho_{i}$, as
$y_i \equiv \rho_{i}/\rho_{\rm th}$.
We assume each detector has identical (arbitrary) sensitivity,
finding:
\begin{equation}\label{eq:scalesnr}
f_{y_i}=\frac{3}{n}\displaystyle\int_{\sum_{j=1}^{N} R_{ij} \ge \frac{1}{y_i^2}} \frac{1}{y_i^4}\; P_i^{3/2} {\rm sin}\,\theta \; {\rm sin}\,\iota  \, d\theta\,d\phi\,d\psi\,d\iota.
\end{equation}
This expression gives the distribution of SNR detected by each
individual detector as part of a given detector network. In Fig.~\ref{fig:scale} we show the 
SNR distribution for each detector within a network composed of two (LIGO-Hanford [H]
and LIGO-Livingston [L]) and 
three (H,L, and Virgo [V]) detectors. 
For any two detector network, the SNR distributions for the individual
detectors will be identical if the detectors operate at the same sensitivities. 
We find that Virgo tends to detect lower SNR
values compared to the LIGO detectors
when operated within the HLV network, even if the
sensitivities of all three instruments are comparable. 
This is because the H and L detector arms are more closely aligned,
and therefore more sources will be detected in the optimal directions for H and
L (``overhead'' for those detectors), leading to weaker SNR in Virgo.

We note that in practice GW searches often use a complicated detection
threshold to better handle the presence of non-Gaussian noise (i.e.,
glitches). To get a sense of the importance of this, we have implemented a
combined coherent/coincident threshold approach, where we demand $\rho_{\rm
  net}>12$ and also implement an individual threshold of $\rho_i>5$ in at least
two detectors. We find that this additional restriction eliminates less than 1\%
of events in the HLV network, and therefore does not substantially impact any of our
predicted SNR distributions.


\section{Discussion}
Bernard Schutz has emphasized that there exists a
  universal distribution of signal-to-noise (SNR) that will be measured for
  gravitational wave sources~\cite{2011CQGra..28l5023S}. This distribution is presented in
  Eq.~\ref{eq:distribution}, and assumes only that the spatial density of the sources does not evolve; it makes
no assumptions about the nature of the sources (e.g., binary coalescence or
supernovae or something else entirely), the properties of the sources (e.g., mass distribution of
binaries, inclination distribution, sky locations), or the properties of the GW network
(e.g., pulsar timing arrays or interferometers of any number, sensitivity, or
location [including ground or space]).
We have derived the universal distribution for the 
  loudest (highest SNR) events, for any given number of detected events. 
When there are $N$ detections above network threshold
$\rho_{\rm th}$, 90\% of the time the loudest event will have SNR larger than $0.76\rho_{\rm
  th} N^{1/3}$. This loudest event may play an important
role in binary parameter estimation, and is expected to be particularly well localized on
the sky, since localization scales roughly as 1/SNR$^2$. If we consider the
first four detections by the Advanced LIGO network (or any sources within any network
with a network threshold of $\rho_{\rm th}=12$), we find that half the time the
loudest event will be louder than $\rho=22$, and the localization area will
shrink by a factor of $\sim3$ compared to threshold events.

Our results are similar to the $V/V_{\rm max}$ test, which is a geometric test
used for electromagnetic astronomical sources~\cite{1968ApJ...151..393S}. For any
population one calculates the volume enclosed to each individual source, $V$, and the
maximum volume to which that source could have been observed, $V_{\rm
  max}$. If there is
no evolution in the source population, simple geometric arguments imply that the
observed values of $V/V_{\rm max}$ must be uniformly distributed between 0 and
1. The same test can be applied to non-evolving GW sources at low redshift:
since SNR scales inversely with $D$, and since volume scales as $D^3$,
we find $ V/V_{\rm max}\sim (D/D_{\rm max})^3\sim(\rho_{\rm th}/\rho)^3$  distributes
uniformly between 0 and 1. We have focused on the SNR
distribution instead, since this quantity is directly measured by GW detectors.
However, the $V/V_{\rm max}$ distribution remains true to arbitrary
redshift (modulo gravitational lensing, which adds noise and
may also introduce magnification bias to all high-$z$ distributions). This is not true for the SNR distributions discussed above, since at
high redshift two additional effects come in: luminosity distance (which sets
the SNR) and comoving
distance (which is relevant for the comoving volume) start to deviate from each
other, and the source redshift affects where the source is found relative to
the frequency response of the GW detectors.  Both of these effects break the
  universality of the SNR distributions.
The latter effect depends upon properties of the source population and detector
noise curves; for binary systems the effect is
encapsulated in the redshift dependence of $I_7$. 
For example, using the Einstein Telescope noise curve~\cite{2008arXiv0810.0604H} 
we find a $\sim 10\%$ suppression from the form in Eq.~\ref{eq:distribution}
for binary neutron stars detected at
$\rho_{\rm th}=12$ (corresponding to a horizon of $z\sim 1.2$).
This effect grows to $25\%$ and $60\%$ as the binary masses increase to
$3\msun$--$3\msun$ and $10\msun$--$10\msun$, respectively (corresponding to horizons of $z\sim
2.8$ and $z\sim 4.8$).
In principle, precise measurements of the
distribution of SNR could be used to infer the intrinsic mass distribution of
binary systems, as well as probe the cosmological parameters by measuring
directly the evolution of the cosmological volume.
In practice the evolution in the rate density of the source populations
dominates over the cosmological effects, and we are more likely to be able to
measure the former than the latter.

These universal distributions are a robust prediction for all GW sources
  and for all GW networks, and therefore serve as an important internal consistency check for the
detectors. For example, in LIGO's 6th and Virgo's 3rd science run
there was a ``blind'' hardware injection event
intended to test the data analysis procedures.  This event is presented in
Fig.~3 of~\cite{2012PhRvD..85h2002A}: the ``false'' coincidence events are found
at SNR below 9.5, while the single injection event stands out at SNR of $\sim 12.5$.
Given our universal distribution, we can calculate the probability of
having a single event at $\rho=12.5$ with no other events down to a threshold of
$\rho=9.5$. Instead of the loudest event, we are now interested in the
``quietest'' event; following the approach in Eq.~\ref{eq:maxsnr}, we find that
the distribution of the lowest SNR for $N$ events is: $f_{\rho_{\rm 
  min}}=3N\, \rho_{\rm th}^{3N}/\rho_{\rm min}^{3N+1}$. For the blind injection
we have $N=1$, and the probability that the first event will
have $\rho\geq12.5$ when the threshold is $\rho_{\rm th}=9.5$ is
$(9.5/12.5)^3=44\%$. We conclude that the injection event was not unlikely, even
in the absence of any other events down to $\rho=9.5$~\footnote{The SNR values
presented in~\cite{2012PhRvD..85h2002A} are based on a ``reweighted'' SNR statistic,
rather than SNR of the form used in this paper. Converting to the
latter leads to values of $\rho\sim24$ and $\rho_{\rm th}\sim12$, leading to a
probability $\sim(1/2)^3=13\%$.}.
The first LIGO/Virgo events must follow statistical expectations from our universal
distributions, and this will be an important sanity check.

In conclusion, all non-evolving low redshift populations found in all GW
detectors must follow the SNR distribution presented in
Eq.~\ref{eq:distribution}. This distribution serves as an important internal
consistency check, and offers the opportunity to test instrumental calibration and
sample completeness, as well as testing for source and cosmological evolution.
In addition, we robustly predict the distribution of the loudest events. These
events {\em must}\/ be found, and will play an important role in gravitational
wave astrophysics.


\begin{acknowledgments}
We acknowledge very valuable discussions with Duncan Brown.
The authors were supported by NSF CAREER grant
PHY-1151836. They were also supported in part by the Kavli Institute for
Cosmological Physics at the University of Chicago through NSF grant PHY-1125897
and an endowment from the Kavli Foundation and its founder Fred Kavli.
In addition, DEH acknowledges the hospitality of the Aspen Center for
Physics, which is supported by NSF grant PHYS-1066293.

\end{acknowledgments}




\bibliography{ref_loudest}

\end{document}